\documentclass[10pt]{article}
\usepackage{amssymb}
\usepackage{fullpage,times,mathptmx}
\usepackage{amsmath,amsthm}
\usepackage{url}

\newtheorem{theorem}{Theorem}
\newtheorem{lemma}{Lemma}
\newtheorem{corollary}{Corollary}
\newtheorem{proposition}{Proposition}
\newtheorem{observation}{Observation}

\newcommand{\gpf}[1]{\textrm{gpf}(#1)}
\theoremstyle{definition} 
\newtheorem{example}{Example}

\title{Abelian Primitive Words}
\author{Michael Domaratzki \\
Department of Computer Science \\
University of Manitoba \\
Winnipeg, MB R3T 2N2 \\
Canada \\
\texttt{mdomarat@cs.umanitoba.ca} \\
\and
Narad Rampersad \\ 
Department of Mathematics \\
University of Li\`ege \\
4000 Li\`ege \\
Belgium \\
\texttt{narad.rampersad@gmail.com}
}
\date{}

\begin{document}

\maketitle

\begin{abstract}
We investigate Abelian primitive words, which are words that are not Abelian 
powers. We show the set of Abelian primitive
words is not context-free. We can determine whether a word is Abelian primitive in 
linear time. Also different from classical primitive words, we find that a word may have 
more than one Abelian 
root. We also consider enumeration of Abelian primitive words. 
\end{abstract}
\section{Introduction}

Repetition in words is a well-studied topic, and many of the results in this
area can be classified into two distinct research areas: the theory 
of formal languages and the study of combinatorics on words. In these two areas,
the focus on repetition is slightly different: in formal language theory, research
focuses on the properties of languages containing words with different types of repetition, 
while in combinatorics on words, research typically concentrates on the existence or 
non-existence of individual words which avoid certain repetitions, and 
combinatorial enumeration of words with or without repetitions.

An example of a long-standing area of research relating to repetition in both the theory 
of formal languages and combinatorics on words are primitive words: a word $x$ is primitive if it cannot be expressed
as a repetition of some shorter word $y$.  In combinatorics on words, an elegant proof 
of the number of primitive words of length is given using M\"obius inversions (see,
e.g., Lothaire \cite{L97}). However, in formal language theory, it is unknown whether the
set of primitive words are a context-free language or not (see, e.g., D\"om\"osi \textit{et al.}~\cite{DHIKK93}).
However, it is known that a closely related set, the set of Lyndon words, is not
context-free \cite{BB97}.

In combinatorics on words, a parallel notion to standard repetition is Abelian 
repetition. A word $x$ is an Abelian power if it can be divided into blocks
$x = x_1 x_2 \cdots x_n$ where every block $x_i$ is a permutation of every other block.

In this paper, we consider the application of Abelian repetition to the concept of
primitivity. Despite the naturalness of this application, the concept does not
appear to have attracted much attention before\footnote{We have found a reference 
to a research project
studying Abelian primitive words on the web at \url{http://bit.ly/9NWqSI}, but have 
been unable to obtain a copy of
any associated works.}. In a related concept, Czeizler \textit{et al.}~\cite{CKS10} 
study repetitions
with only limited rearrangement. We study the language of Abelian primitive words, a formal
language theoretic question, as well as the number of Abelian primitive words of
a given length, a problem in combinatorics on words. 

\section{Definitions}

For additional background in formal languages and automata theory, see Rozenberg
and A.~Salomaa \cite{RS97}. Let $\Sigma$ be a finite set of letters, called an 
alphabet. 
A string over $\Sigma$ is any finite sequence of letters from $\Sigma$. 
The string containing no symbols, the empty string, is denoted $\epsilon$. The set $\Sigma^*$ is the
set of all strings over $\Sigma$. A language $L$ is any subset of $\Sigma^*$.
If $x = a_1 a_2 \cdots a_n$ is a string, with $a_i \in \Sigma$, then the length of
$x$, denoted by $|x|$, is $n$. For $a \in \Sigma$ and $w \in \Sigma$, $|w|_a$ is the number of 
occurrences of $a$ in $w$.

For languages $L_1,L_2 \subseteq \Sigma^*$ the left quotient of 
$L_1$ by $L_2$, denoted $L_2 \setminus L_1$, is defined by 
\[ L_2 \setminus L_1 = \{ x \in \Sigma^* \ : \ \exists y \in L_2 \textrm{ such that }  
yx \in L_1 \}. \]

Given an (ordered) alphabet $\Sigma = \{ a_1, \dots, a_n \}$, the Parikh vector of a 
word $w \in \Sigma^*$ is
$\Psi(w) = ( |w|_{a_1} , |w|_{a_2} , \dots, |w|_{a_n} )$.  For the alphabet $\Sigma = \{a,b\}$, we assume $a<b$.
Thus, for example $\Psi(abbab) = (2,3)$.

We first recall the standard notions of primitive words. A word $w$ is primitive if $w$ cannot be written as
$z^k$ for $z \in \Sigma^*$ and $k \ge 2$. If $w$ is not primitive, then there is a unique primitive word $u$ 
such that $w = u^k$ for some $k \ge 2$.  For an alphabet $\Sigma$, the set of all primitive words $w \in \Sigma^*$  
is denoted $Q(\Sigma)$ or simply $Q$ if $\Sigma$ is understood.

We now turn to the generalization of these notions to Abelian repetitions.
A word $w$ is a $n$-th Abelian power if $w = u_1 u_2 \cdots u_n$ for some $u_1,u_2,\dots,u_n$ such that for all
$1 \le i,j \le n$, $\Psi(u_i) = \Psi(u_j)$. That is, each $u_j$ with $j \ge 2$ is a permutation of $u_1$.

We say that a word $w$ is Abelian primitive (or A-primitive, for short) if $w$ fails to
be a $k$-th Abelian power for
every $k \ge 2$.  For an alphabet $\Sigma$, the set of all A-primitive words $w \in \Sigma^*$ is denoted by 
$AQ(\Sigma)$ or simply $AQ$ if $\Sigma$ is understood.

\begin{example}
The word $w = aabbab$ is A-primitive, while $u = aabbabab$ is not, as $u = xy$ where $\Psi(x) = \Psi(y) = (2,2)$.
\end{example}

Let $w$ be an Abelian power. Then we say that an word $u$ is an Abelian root 
(or A-root) of $w$ if $w = u u_1 u_2 \cdots u_n$
for some $u_1,\dots,u_n \in \Sigma^*$ with $\Psi(u) = \Psi(u_i)$ for all $1 \le i \le n$. 
If $w$ has an A-root $u$ which is also A-primitive, then we say that $u$ is an A-primitive
root of $w$. Two A-primitive roots $u,v$ 
of a word $w$ are distinct if $|u|$ does not divide $|v|$ or vice versa.
On the other hand, we note the following simple but useful fact:

\begin{observation}\label{obv_ob}
If a word $x$ has an A-root of length $k$, then $x$ also has an A-root of length $k'$ for all $k'$ where $k$ divides $k'$ and $k'$ divides $n$. 
\end{observation}

We recall some notation from number theory.  
Recall that if $r,z$ are integers, $r \mid z$ denotes that
$r$ divides $z$, i.e., $z = rk$ for some $k \ge 0$.
We say that a set of integers $S$ is division-free if 
$x \nmid y$ and $y \nmid x$ for all $x,y \in S$. 
For all $n \ge 2$, let $\omega(n)$ denote the number of prime divisors of $n$,
while $\omega'(n)$ is the number of prime divisors of $n$ with 
multiplicity\footnote{The notation $\Omega(n)$ is also used for
what we call $\omega'(n)$, but we reserve $\Omega$ for denoting asymptotic function growth.
Our notation is from Bach and Shallit \cite{BS97}.}. 
Thus, if $n\ge 2$ and $n = p_1^{\alpha_1} 
p_2^{\alpha_2} \cdots p_k^{\alpha_k}$ is its prime factorization, then 
$\omega(n) = k$ and $\omega'(n) = \sum_{i=1}^{k} \alpha_i$. 
We also let $d(n)$ be the number of divisors of $n$, i.e., $d(n) = \prod_{i=1}^{k}(1 + \alpha_i)$.

\section{Non-context-freeness of $AQ$}

We now show that the set $AQ$ of all A-primitive words is not context-free. This is
in contrast to the set of ordinary primitive words $Q$, for which it is unknown whether
they are a context-free language or not. We begin with two preliminary propositions.

\begin{proposition}\label{prim_prop}
Let $p$ be a prime and $x = aabb(ab)^{p-2}$. Then $x$ is A-primitive.
\end{proposition}

\begin{proof}
Note that $|x| = 2p$.  If $x$ is not A-primitive, then one of 
three cases occurs:
\begin{enumerate}
\item[(a)] $x = u^{2p}$ for some letter $u$, 
\item[(b)] $x = u_1 u_2 \cdots u_p$ for 
words $u_1,\dots,u_p$ of length two, or 
\item[(c)] $x = v_1 v_2$ for words $v_1,v_2$ of length $p$.
\end{enumerate}

The first of these possibilities cannot occur, as $x$ contains occurrences of both 
$a$ and $b$. 
The second case is also not possible, since if so, we would have $u_1 = aa$ and $u_2 =bb$,
which do not have matching Parikh vectors. Thus, we must have that $x = v_1v_2$
for $|v_1|=|v_2|=p$. We have three subcases:
\begin{enumerate}
\item[(a)] if $p=2$, then we are in the previous case, i.e., $v_1 = aa$.
\item[(b)] if $p=3$, then $v_1 = aab$ and $v_2 = bab$.
\item[(c)] otherwise $p>3$ and $v_1 = aabb(ab)^{(p-5)/2}a$ which has Parikh
vector $( (p-5)/2 +3, (p-5)/2+2)$, and $v_2 = b(ab)^{(p-1)/2}$ which has
Parikh vector $( (p-1)/2, (p-1)/2 + 1)$. We can see that the number of occurrences
of $a$ in $v_1$ is even, while in $v_2$ it is odd or vice versa.
\end{enumerate}
\end{proof}

\begin{proposition}
Let $M = AQ \cap aabb(ab)^*$. Then 
\[ M = \{ aabb(ab)^{p-2} \ : \ p \textrm{ is prime. } \}. \]
\end{proposition}

\begin{proof}
The right-to-left inclusion is immediate from Proposition~\ref{prim_prop}.

For the reverse inclusion, let $x \in M$. Then $|x| = 2n$ for some $n \ge 2$. Suppose, 
contrary to what we want to prove, that $x$ is not of the form $aabb (ab)^{p-2}$ for 
some prime $p$.  Then we must have that $n$ is not prime. Let $q$ be a prime factor 
of $n$ and note
that 
\[ x  =  (aabb (ab)^{q-2}) \cdot ((ab)^{q})^{n/q-1} \] 
and that all factors of length $2q$ have $q$ occurrences of $a$ and $q$ occurrences
of $b$.  Further, $aabb (ab)^{q-2}$ is an A-primitive root by Proposition~\ref{prim_prop}. 
\end{proof}

We can now show that the set of all $A$-primitive words is not context-free.

\begin{theorem}
The set $AQ$ is not context-free.
\end{theorem}

\begin{proof}
We prove that $M$ is not context-free. Let $M' = h^{-1}(\{ aabb \}\setminus M)$
where $h : \{a \}^* \to \{a,b\}^*$ is the morphism $h(a) =ab$. Then
$M' = \{ a^{p-2} \ : \ p \textrm{ is prime } \}$. As the context-free languages are closed
under quotient by regular sets and inverse homomorphism, $M'$ is context-free if
$M$ is. But as $M'$ is unary, if it is a context-free language then it is
also regular. But by the pumping lemma, we can see that $M'$ is not regular. 
Thus, neither are $M$ or $AQ$.
\end{proof} 

The set of all non-trivial Abelian powers, $\overline{AQ}$, 
is also non-context-free, as can be seen through, e.g., the intersection
\[ \overline{AQ} \cap  a^*b a^*b a^*b  = \{ a^n b a^n b a^n b \ : \ n \ge 0 \}. \]
For discussion on the complexity of the language of marked Abelian squares 
and its relation to iterated shuffle and deletion operations, see
Domaratzki \cite[Sect.~8.4.1]{D04-thesis} and J\c{e}drezejowicz and Szepietowski 
\cite[Ex.~3.2]{JS01}. Using the interchange lemma, Gabarr\'o \cite{G85} has proven 
that the language 
$\{ u w_1 w_2 v \ : \ u,w_1,w_2,v \in \Sigma^*, \Psi(w_1) = \Psi(w_2) \}$ of
words containing an Abelian square is not context-free.

\section{Complexity of $AQ$}

Through an elegant pattern matching algorithm \cite[Thm.~13]{P94}, it is known that we can determine whether a word is primitive in linear time. We now consider this problem for 
A-primitive words. Throughout this section, we consider the size of the alphabet to be
a fixed constant.  In order to illustrate the basic principles of the algorithm, we 
begin with an $O(n \log n/\log \log n)$ algorithm: 

\begin{quote}
\begin{verbatim}
def isAbelPrim(w):
  n = len(w)
  if n==1:
    return True
  PF = { p : p is prime, p | n } 
  D = { n/p : p in PF }
  for d in D:
    if w has an A-root of length d:
      return False
  return True
\end{verbatim}
\end{quote}

Suppose that $w \in AQ$. Then $w$ certainly does not have a A-root whose
length is any the periods in $D$, thus \verb@isAbelPrim@ returns true.  On the 
other hand, if 
$w \notin AQ$ with $|w| > 1$, then $w$ has an A-root of length 
$r$ for some $r \mid n$ with $r < n$.   
There exists $d_r \in D$ such that $r\mid d_r$ ($r$ may also divide other $d \in D$, but it 
is enough to know it divides some $d_r$).  By Observation~\ref{obv_ob}, on the loop 
of \verb@isAbelPrim@ with $d = d_r$, the algorithm will return false.  

One iteration of the loop in \verb@isAbelPrim@ will take time $O(n)$, by walking across 
$w$ and computing the Parikh vectors for each block of length $d \in D$.  Thus, the 
runtime of the algorithm is $O(p(n) + n \omega(n))$ where $p(n)$ is the time required to 
calculate the set \verb@PF@. 

We claim that even using trial division (rather than more complex methods such as, 
e.g., general number field sieve \cite{CP05}), we have $p(n) \in O(\sqrt{n} \log n)$. 
Consider the following algorithm:

\begin{quote}
\begin{verbatim}
def PF (n):
  pf = []
  while True:
    p = 2
    found = False
    while ( p <= math.ceil(math.sqrt(n)) and (not found)):
      if (n % p == 0):
        found = True
        pf.append(p)
        while (n % p == 0):
           n /=  p 
      p += 1
    if (not found):
      break
  if (n != 1):
    pf.append(n)
  return pf
\end{verbatim} 
\end{quote}

The method \verb@PF@ calculates the prime factors of $n$ by repeatedly finding the
least prime $p$ dividing $n$ and factoring out the largest power of $p^{\alpha}$ 
which divides $n$.  Then this process is repeated on $n/p^{\alpha}$. 

As for the running time of \verb@PF@, let $n = \prod_{i=1}^{k} p_i^{\alpha_i}$ be the
prime factorization of $n$. The outer while loop executes $\omega(n) =k$ times,
once for each $p_i$ dividing $n$,
while one execution of the inner two while loops takes $O(\sqrt{n} + \alpha_i)$ time. 
Thus, the total run-time is $O(\sum_{p_i \mid n } \sqrt{n} + \alpha_i) = O(\sqrt{n} \omega(n)
+ \omega'(n))$. As $\omega(n) \in O(\log n/ \log \log n)$ \cite[Thm.~8.8.10]{BS97} and 
$\omega'(n) \in O(\log n)$ \cite[Sect.~22.10]{HW00}, this gives the claimed worst 
case running time for \verb@PF@. 

%
%

Thus, the running time of \verb@isAbelPrim@ is $O(n \omega(n))$. Using the same
estimate on the worst-case growth of $\omega(n)$, we obtain the following result:
 

\begin{theorem}
Given $x$, there is an algorithm to determine if $x \in AQ$ which runs in time 
$O(n \frac{\log n}{\log \log n})$ time in the worst case. 
\end{theorem}

For space complexity, we briefly note that the set $AQ$ is in 
$\textsc{dspace}(\log(n))$. To see
this, if we are testing whether a word is of the form $u_1 u_2 \cdots u_n$ where
$\Psi(u_i) = \Psi(u_j)$ for all $i,j$, we can use log-sized counters to keep track of 
the current prefix length $|u_1|$, block number $j$ ($2 \le j \le n)$ and the
values of the Parikh vectors for $u_1$ and $u_j$. Viewing the alphabet size as
constant, this is a constant number of counters.

\subsection{A linear time algorithm for recognizing $AQ$}

We can improve the algorithm \verb@isAbelPrim@ from the previous section by caching 
commonly used Parikh vectors, and obtain a linear time algorithm. Let $\gpf{n}$ be the 
greatest prime factor of $n$.  Then we note that if $\gpf{n}^2\mid n$, every $d \in D$ is 
divisible by $\gpf{n}$, while if $\gpf{n}^2 \nmid  n$, then every $d \in D$ is divisible by 
$\gpf{n}$ except $d = n/\gpf{n}$.  In both cases, we will precompute the Parikh vectors of 
length $\gpf{n}$ in order to compute the Parikh vectors of length $d$ for all 
$d \in D$ which are divisible by $\gpf{n}$.

Let $w$ be our input word of length $n$ and write $w = w_1 w_2 \cdots w_{n/\gpf{n}}$ where each block 
has length $\gpf{n}$. Let $\mathbf{u}_i = \Psi(w_i)$ for $1 \le i \le n/\gpf{n}$.  
Note then that if $\gpf{n}\mid d$, then the blocks of $w$ of length $d$ have Parikh 
vectors of the form 
\[ \sum_{j=1}^{d/\gpf{n}} \mathbf{u}_{kd/\gpf{n} + j} \] 
for some $1 \le k < \gpf{n}$. 
Thus, we can compute these Parikh vectors quickly by summing the precomputed 
$\mathbf{u}_i$. 

\begin{quote}
\begin{verbatim}
def isAbelPrimLin(w):
  n = len(w)
  PF = { p : p is prime, p|n }
  gpf = max(PF)
  D = { n/p : p in PF }
  if ( n % (gpf**2) != 0):
    D.remove(n/gpf)
    calculate Parikh vectors of length n/gpf.
    if w has an A-root of length n/gpf:
       return False
  for i in range(0,n/gpf(n)):
    u[i] = Parikh(w,i,gpf(n))
  for d in D:
    calculate Parikh vectors of length d (using u[i])
    if w has an A-root of length d:
      return False
  return True
\end{verbatim}
\end{quote}

Here, we let \verb@Parikh(w,i,j)@ be a method which computes the \verb@i@-th Parikh vector 
of length \verb@j@ in the word \verb@w@.

This modified implementation has the same correctness as the previous implementation, 
as the same tests are performed.  We now show the claimed $O(n)$ run time.  
Computing \verb@D@ and \verb@PF@ is the same as in  
\verb@isAbelPrim@ and can be done in linear time. 
Having computed \verb@PF@, the 
calculation of $\gpf{n}$ takes $O(\omega(n)) = O(\log n/\log \log n)$ time.  
In the case where 
$\gpf{n}^2 \nmid n$, the time to execute the additional statements is $O(n)$ time.  
Similarly, the computation of the Parikh vectors \verb@u[i]@ takes time $O(n)$. 

Consider the execution of the final for loop. For $d \in D$, we need $O(d/\gpf{n})$ time 
to compute one Parikh vector of a subword of $w$ of length $d$, so to compute all 
$n/d$ such vectors requires 
time $O(n/\gpf{n})$.  To test the equalities of all these $n/d$ vectors (implied by the if 
statement) requires time $O(n/d) = O(p)$ where $d=n/p$.  Thus, the worst case 
running time of the loop is 
\[ \sum_{p\mid n} \left( O(\frac{n}{\gpf{n}}) + O(p) \right) = O(n \frac{\omega(n)}{\gpf{n}}) 
+ O( \sum_{p\mid n} p ). \]
We now estimate the first quantity.

\begin{proposition}
For all integers $n$, $\omega(n)/\gpf{n} \le 2/3$.
\end{proposition}

\begin{proof}
Note that if $\omega(n) = r$ for some integer $r$, then $\gpf{n} \ge p_r$ (where $p_r$ 
is the $r$-th prime), since if $n$ has $r$ prime factors, the minimum possible value for 
its largest prime factor is $p_r$. A simple induction proves that $p_r > 2r-1$ for 
$r \ge 2$. Thus, $\omega(n)/\gpf{n}$ is maximized by $x/(2x-1)$ for all $n$ with at 
least two prime factors. But $\frac{x}{2x-1}$ is maximized at $n =2$ on the 
interval $n \ge 2$. Thus, $\omega(n)/\gpf{n} \le 2/3$ for all $n$ with at most two prime 
factors. For prime powers, $\omega(n)/\gpf{n} \le 1/n < 2/3$. 
\end{proof}

Finally, we have that $\sum_{p\mid n} p \le n$. Thus, the total running time of the loop is 
$O(n)$. 

\begin{theorem}
Given $x$, there is an algorithm to determine if $x \in AQ$ which runs in time 
$O(n)$ time in the worst case. 
\end{theorem}

\section{Words with multiple A-primitive roots}

We show that unlike classical primitive words, a word may have multiple 
distinct A-primitive roots. This fact was essentially noted by Constantinescu and Ilie \cite{CI06}
who constructed an infinite word $w$ with two distinct Abelian periods. We generalize this 
to show that for all $n \ge 2$, we can construct a word with $n$ distinct A-primitive roots. 

For all $n \ge 1$, let $Q_n = 2 \cdot \prod_{i=1}^{n} p_i$, where $p_i$ is the $i$-th
prime for $i \ge 1$, with $p_1 = 2$.  Then for all $n \ge 1$, let $w_n$ be the word
defined by 
\[ w_n = aabb (ab)^{(Q_n-4)/2}. \]
Note that $|w_n| = Q_n$. For example,
\[ w_2  =  aabbabababab.  \]

\begin{lemma}
For all $n \ge 2$, the word $w_n$ has $n$ distinct A-primitive roots. In
particular, the words 
\[ r_m = aabb(ab)^{p_m-2} \]
for all $n \ge m \ge 1$ are A-primitive roots of $w_n$.
\end{lemma}

\begin{proof}
First, note that $ab$ is not an A-primitive root of $w_n$, as it is not a
prefix of $w_n$.  

Let $1 \le m \le n$. Then the first subword of $w_n$ of length $2p_m$ is
$r_m$. All subsequent subwords of $w_n$ of length $2p_m$ are $(ab)^{p_m}$.
All subwords have Parikh vector $(p_m,p_m)$. 

Finally, note that the lengths of $r_m$ form a division-free set 
$\{ 2p_n \ : \ 2 \le m \le n \}.$ Thus, all $r_m$ are A-primitive roots of $w_n$.
\end{proof}

The following lemma shows that a word may not have A-primitive roots whose
lengths are coprime.

\begin{lemma}
If $w$ has two distinct A-primitive roots $u,v$ where $|u| = \ell_1$, $|v|=\ell_2$, 
then $\gcd(\ell_1,\ell_2) \ge 2$.
\end{lemma}

\begin{proof}

Assume that $w$ has two distinct A-primitive roots as above: 
$w = u_1 u_2 \cdots u_m$ and $w = v_1 v_2 \cdots
v_n$ where $|u_i| = \ell_1$, $|v_j|=\ell_2$. Assume, contrary to what we want to prove
that $\gcd(\ell_1,\ell_2) = 1$. 

First note that $m \ge \ell_2$. To see this, note that $|w|= m\ell_1 = n\ell_2$ and we 
have that $\ell_2\mid \ell_1m$. If $m < \ell_2$, and as $\ell_1$ and $\ell_2$ are coprime, $\ell_2\mid \ell_1m$ is a contradiction.

Thus $m \ge \ell_2$ and $n \ge \ell_1$ as well. As $\gcd(\ell_1,\ell_2)=1$,
there exist $r,s \ge 0$ such that $r\ell_1 = s\ell_2-1$ (or $r\ell_1=s\ell_2+1$, which is
proven similarly).  As $m \ge \ell_2$ and $n \ge \ell_1$, we can assume that $s \le n$ and
$r \le m$. 

Thus, the prefix $v' = v_1 v_2 \cdots v_s$ of $w$ of length $s\ell_2$ is one
letter longer than the prefix $u'= u_1 u_2 \cdots u_r$.  Without loss of 
generality, let $a$ be the last symbol of $v_s$, which is also the first symbol 
of $u_{r+1}$.  Let $\alpha = |u_1|_a$ and $\beta = |v_1|_a$. 
Counting the occurrences of $a$ in $u'$ and $v'$, we get 
\begin{equation}\label{alpha_1}
 r\alpha = s \beta -1. 
\end{equation}
Now consider that the prefix of $w$ of length $\ell_1\ell_2$ is
$u_1 \cdots u_{\ell_2} = v_1 \cdots v_{\ell_1}$. 
Considering $v'' = v_{s+1} \cdots v_{\ell_1}$ and $u'' = u_{r+1} \cdots u_{\ell_2}$, and 
again counting the occurrences of $a$, we also have
\begin{equation}\label{alpha_2}
(\ell_2-r) \alpha = (\ell_1-s) \beta + 1. 
\end{equation}
Equating both (\ref{alpha_1}) and (\ref{alpha_2}) in terms of $\alpha$, we get 
\[  r ( (\ell_1-s) \beta + 1) = (\ell_2 - r)  (s\beta - 1). \]
Solving for $\beta$ gives $\beta = \ell_2$. Thus, we have that $v_1 \in a^+$ and
thus $w$ only has $A$-primitive root $a$, a contradiction. Thus,
$\gcd(\ell_1,\ell_2) \ge 2$.
\end{proof}

\section{Number of Abelian Primitive Roots}

We now turn to the number of A-primitive roots a word may have, as a function
of its length. As shown in the previous section, for any $n$, we can construct a word 
with $n$ A-primitive roots. In this section, we improve this to give a tight bound on 
the number of A-primitive roots a word may have. 

\subsection{Upper Bound}
We first give an upper bound on the number of A-primitive roots a word may have.
%
We need an estimate $d(n)$ \cite{BS97}.

\begin{theorem}\label{bs_1}
The function $d(n)$ satisfies $d(n) \in O(2^{\log n / \log \log n})$.
\end{theorem}

We will also use a result by de Bruijn \textit{et al.} \cite{BEK51} (see also Anderson \cite{A02}):

\begin{theorem}\label{deB_1}
Let $n = p_1^{\alpha_1} p_2^{\alpha_2} \cdots p_k^{\alpha_k}$ be the prime factorization
of $n \ge 2$. Let $D(n)$ be the set of integers defined by  
\[ D(n) = \{ p_1^{\beta_1} p_2^{\beta_2} \cdots p_k^{\beta_k} \ : \ 
\forall i ( \beta_i \le \alpha_i) \textrm { and } 
\sum_{i=1}^{k} \beta_i = \lfloor \omega'(n)/2 \rfloor \}. \]
Then $D(n)$ is a maximal anti-chain in the divisor lattice of $n$.
\end{theorem}

In other words, $D(n)$ is the largest division-free set of divisors of $n$.
Anderson \cite{A02} gives the following estimate on the size of $D(n)$, which
we denote $s(n)$:

\begin{theorem}\label{anderson}
Let $n = p_1^{\alpha_1} p_2^{\alpha_2} \cdots p_k^{\alpha_k}$ be the prime factorization
of $n \ge 2$. Let $A(n) = \frac{1}{3} \sum_{i=1}^{k} \alpha_i (\alpha_i+2)$. Then the
maximal anti-chain in the divisor lattice of $n$ has size $s(n) = \Theta( d(n) / \sqrt{A(n)})$.
\end{theorem}

Now a word $w$ of length $n$ has at most $|D(n)|$ A-primitive roots: if $r$ is an
A-primitive root, then $|r|$ divides $|w|$ and $|r|$ is not divisible by the length of any 
other A-primitive root. Thus, we can obtain the following result:

\begin{theorem}
If $w$ is a word of length $n$, the number of distinct A-primitive roots is 
$s(n) \in o(2^{\log n / \log \log n})$.
\end{theorem}

\begin{proof}
By Theorem~\ref{deB_1}, if $w$ is a word of length $n$, then $w$ has at most
$s(n)$ distinct A-primitive roots. 
By Theorem~\ref{bs_1} and Theorem~\ref{anderson}, 
$d(n) \in o(2^{log n / \log \log (n)})$.  
Thus, the result follows.
\end{proof}

We can use a result of Anderson \cite{A67} which gives the average order of $s(n)$: 

\begin{theorem}
As $\omega'(n) \to \infty$, we have 
\[ s(n) \le \left ( \sqrt { \frac{2}{\pi} + o(1) }  \right ) \frac{ d(n)}{\sqrt{ \omega'(n) } } . \]
As $n \to \infty$, 
\[ \sum_{m \le n} \frac{ d(m)}{\sqrt{ \omega'(m) } } \sim \frac{n \log n}{ \sqrt{2 \log \log n}}. \]
\end{theorem}

\subsection{Lower Bound}

For a lower bound on the number of A-primitive roots a word may have, we give 
an explicit construction. For any $n \ge 2$, 
let $T(n) = \{ kd \ : \ k \in \mathbb{N}, d \in D(n), kd \le n \}$. Let
$t_1 < t_2 < \cdots < t_{m_n} = n $ be the $m_n$ elements of $T(n)$ in sorted order. Define
\[z_n =  a^{t_1} b^{t_1} \prod_{i=2}^{m_n} a^{t_i-t_{i-1}} b^{t_i - t_{i-1}}.\]
Note that $z_n$ is a word of length $2n$ with $\Psi(z_n) = (n,n)$.

\begin{example}
If $n = 30$, then $D(30) = \{ 2,3,5 \}$. In this case 
\[ T(30) = \{ 2, 3, 4, 5, 6, 8, 9, 10, 12, 14, 15, 16, 18, 20, 21, 22, 24, 25, 26, 27, 28, 30 \}. \]
With this, we have
\[ z_n = aabbababababaabbababaabbaabbababaabbaabbababaabbababababaabb. \]
\end{example}

\begin{lemma}
Let $n \ge 2$ and $t \in D(n)$. Then $z_n$ has an $A$-primitive root of length $2t$.
\end{lemma}

\begin{proof}
Let $1 \le j \le m_n$ be the index such that $t = t_j$. As
 $t \in D(n) \subset T(n)$, we have that the prefix of $z_n$ of length $2t$ is 
\[ w_n = a^{t_1} b^{t_1} a^{t_2 - t_1} b^{t_2 - t_1} \cdots a^{t - t_{j-1}} b^{t - t_{j-1}}. \]
Note that $\Psi(w_n) = (t,t)$. Now, each additional block of length $2t$ from has the form 
\[ a^{t_{\alpha}} b^{t_{\alpha}} \cdots a^{t_{\beta}} b^{t_{\beta}} \]
for some $\alpha,\beta$ which are differences of successive $t_i$. To see this, note that these
factors of $z_n$ begin and end at positions which are multiples of $t \in D(n)$, so each of the 
breakpoints are elements of $T(n)$.  By telescoping, each of these factors has Parikh vector
$(t,t)$. Thus, $z_n$ is a $n/t$-th A-power. 

Further, $w_n$ must be an A-primitive root of $z_n$. Otherwise, there is some $z \in T(n)$ such that
$z\mid t$, but in this case, $z$ is divisible by some element in $D(n)$, by definition of $T(n)$.
But this gives a contradiction, since $t \in D(n)$ and $D(n)$ is an anti-chain of divisors. 
\end{proof}

\begin{corollary}
For all $n \ge 2$, there exists a word of length $2n$ with $s(n)$ distinct A-primitive
roots.
\end{corollary}

\section{Counting Abelian Primitive Words}

Let $\psi_k(n)$ be the number of primitive words of length $n$ over a $k$-letter alphabet, 
$\psi_k^A(n)$ be the number of A-primitive words of length $n$ over a $k$-letter alphabet and
$\Delta_k(n) = \psi_k(n) -\psi_k^A(n)$. 
Note that $\Delta_k(n) \ge 0$ for all $n$, but we can observe,
e.g., that $\Delta_k(p) = 0$ for all primes $p$. 
Small values of $\psi_k^A(n)$ are given in Figure~\ref{fig__0}.

\begin{figure}[ht]
\begin{center}
\begin{tabular}{|r|r|r|r|r|} \hline
$\downarrow n \quad k \rightarrow$ & 2 & 3 & 4 & 5 \\  \hline
1 & 2 & 3  & 4  & 5 \\  \hline
2 & 2 & 6  & 12 & 20 \\ \hline
3 & 6 & 24 & 60 & 120 \\ \hline
4 & 10 & 66 & 228 & 580 \\ \hline
5 & 30  & 240 & 1020 & 3120 \\ \hline
6 & 36 & 612 & 3792 & 15000 \\ \hline
7 & 126 & 2184 & 16380 &  78120 \\ \hline
8 & 186 & 5922 & 62820 & 382740 \\ \hline
9 & 456 & 19302 & 260952 & 1950420 \\ \hline
10& 740 & 54300 & 1016880  & 9637400 \\ \hline
11& 2046 & 177144 & 4194300   & 48828120 \\ \hline
12& 2972 & 490488 &  16354320& 241776440\\ \hline
13&  8190 & 1594320 & 67108860 & 1220703120\\ \hline
14& 12824 & 4509750 & 263483136 & \\ \hline
15& 30030 & 14227920 & & \\ \hline
16& 52666 & 40888962 & & \\ \hline
17& 131070 & 129140160 & 17179869180 & 762939453120\\ \hline
18& 202392 & 368252856 & & \\ \hline
19& 524286 & 1162261464 & 274877906940 & 19073486328120\\ \hline
20& 859180 & & & \\ \hline
\end{tabular}
\end{center}
\caption{\label{fig__0} Number of A-primitive words $\psi_k^A(n)$ by length ($n$) and alphabet size ($k$).}
\end{figure}

The function $\psi_k(n)$ is well-known (see, e.g., Lothaire \cite{L97}). The formula 
\[ \psi_k(n) = \sum_{d\mid n} \mu(d) k^{n/d} \]
expresses $\psi_k$ in terms of the M\"obius function $\mu$ defined by $\mu(1) = 1$,
$\mu(n) = (-1)^k$ if $n$ is a product of $k$ distinct primes and $\mu(n) =0$ if $p^2 \mid  n$ for some prime $p$.

We can characterize $\Delta_k$ for prime powers exactly: 
\begin{lemma}
For all primes $p$ and all $r \ge 2$,
\[ \Delta_k(p^r) = \sum_{n_1 + n_2 + \cdots + n_k = p^{r-1} } 
{p^{r-1} \choose {n_1 \ n_2 \dots n_k} }
\left ( {p^{r-1} \choose {n_1 \ n_2 \dots n_k} }^{p-1} - 1 \right ).
\]
Here, the sum is taken over all partitions $n_1 + n_2 + \cdots + n_k$ of
$p^{r-1}$.
\end{lemma}

\begin{proof}
Let $x \in Q - AQ$ of length $p^r$.  As $x$ is not A-primitive, it has a A-primitive root of length $p^i$ for some $1 \le i
< r$.  But then $x$ can also be written as 
$x = x_1 x_2 \cdots x_p$ where $|x_i| = p^{r-1}$ and $\Psi(x_i) = \Psi(x_j)$ for all $1 \le i,j \le p$.
Thus, it suffices to count only those $x$ of this form. 

Consider that there are ${p^{r-1} \choose {n_1 \ n_2 \dots n_k} }$ different words $x_1$ of length $p^{r-1}$ 
such that $\Psi(x_1) = (n_1,n_2,\dots,n_k)$ for each partition $n_1 + n_2 + \cdots + n_k = n$.
As recently noted by Richmond and Shallit \cite{RS09}, for a fixed choice of $x_1$, the remainder of the words
$x_2,\dots,x_p$ must satisfy $\Psi(x_j) = \Psi(x_1)$, which can be done in 
${p^{r-1} \choose {n_1 \ n_2 \dots n_k} }$
ways for each $2 \le j \le p$.  Thus, we get a total of 
${p^{r-1} \choose {n_1 \ n_2 \dots n_k} }^{p-1}$ possibilities, and we must exclude the choice
$x_1 = x_2 = x_3 = \cdots = x_p$, as this word is not primitive. 

Thus, multiplying the number of choices of the word $x_1$ and the words $x_2, \cdots, x_p$ and summing over all 
possible Parikh vectors, we get the result.
\end{proof}

The problem of giving a closed form of $\Delta_k(n) $ or $\psi_k^{A}(n)$ for all values of
$n$ is still open.


\section{Equivalence Relations on A-primitive words}

In this section, we consider classical results such as the Lyndon-Sch\"utzenberger Theorem
for classical words in the context of Abelian primitivity.  To do so, we define an
appropriate equivalence relations to replace equality.  

We first note that the A-primitive words are not closed under conjugation. For
example, note that $bbababaa \in AQ$ but $aabbabab \notin AQ$.  Because of this, the 
concept of a Lyndon-type word for A-primitive words is not a straight forward definition
(recall that a primitive word $w$ is a Lyndon word if it is the lexicographically least
word in its class of conjugates).

For all $n \ge 1$, let $\sim_n$ be the binary relation defined on words by $u \sim_n x$ if we can write $u = \alpha_1 \alpha_2 \cdots \alpha_m$ and $x = \beta_1 \beta_2 \cdots \beta_m$ where
\begin{enumerate}
\item[(a)] for all $1 \le i \le m$, $|\alpha_i| = |\beta_i| = n$. 
\item[(b)] for all $1 \le i,j \le m$, $\Psi(\alpha_i) = \Psi(\beta_j)$.
\end{enumerate}

Thus, $\sim_n$ represents that two words can be broken into blocks of length $n$, all of which have the same image under $\Psi$. 

\begin{example}\label{example_code_m1}
Let $n=3$. Then $abc \, acb \, abc \sim_3 cba \, bca \, bca$ as each block $b$ of length
three in both words satisfies $\Psi(b) = (1,1,1)$.
\end{example}

We use $\sim_n$ to investigate relationships with the theory of
codes in the context of commutation.

\begin{theorem}\label{theorem_code_0}
For all words $u,x \in \Sigma^*$,  $ux \sim_n xu$ if and only if there exists $r \ge 1$, $\alpha_1,\dots,\alpha_r,\beta_1,\dots,\beta_r \in \Sigma^*$ such that 
\begin{enumerate}
\item[(a)] for all $1 \le i \le r$, $|\alpha_i \beta_i| = n$.
\item[(b)] for all $1 \le i,j \le r$, $\Psi(\alpha_i) = \Psi(\alpha_j)$ and $\Psi(\beta_i) = \Psi(\beta_j)$.
\item[(c)] there exists $1 \le s < r$ such that $u = \alpha_1 \beta_1 \cdots \alpha_{s-1} \beta_{s-1} \alpha_s$ and
$x = \beta_s \alpha_{s+1} \beta_{s+1} \cdots \alpha_{r} \beta_{r}$.
\end{enumerate}
\end{theorem}

\begin{proof}
($\Leftarrow$) Let $u$,$x$ satisfy the conditions. Then we have that
\begin{eqnarray*}
ux & = & \alpha_1 \beta_1 \cdots \alpha_r \beta_r \\
xu & = & (\beta_s \alpha_{s+1}) (\beta_{s+1} \alpha_{s+2})  \cdots (\beta_{r-1} \alpha_{r}) (\beta_r \alpha_1) (\beta_1 \alpha_2) \cdots  (\beta_{s-2} \alpha_{s-1}) (\beta_{s-1} \alpha_s)
\end{eqnarray*}
Thus, note that with the parenthesization above, we have that each subword in $ux$ of length $n$ has the form  
$\alpha_i \beta_i$ subwords while in $xu$, they have the form $\beta_{i} \alpha_{i+1 (\textrm{mod } r)}$.

Now note that for any value of $i$ and $j$, we have
\[ \Psi (\alpha_i \beta_i) = \Psi(\alpha_i) + \Psi(\beta_i) = \Psi(\alpha_{j+1 (\textrm{mod } r)}) + \Psi(\beta_j) = \Psi(\beta_j \alpha_{j+1 (\textrm{mod } r)}). \] 
Thus, $ux \sim_n xu$. 

($\Rightarrow$) Let $ux \sim_n xu$. Then we can write 
$ux = \gamma_1 \gamma_2 \cdots \gamma_t$ and 
$xu = \eta_1 \eta_2 \cdots \eta_t$ where for all $1 \le i,j \le t$, we have
$\Psi(\gamma_i) = \Psi(\eta_j)$ and $|\gamma_i|=|\eta_j| = n$.  

Assume without loss of generality that $|u|>|x|$. Let $1 \le p <t$ be such that 
\begin{eqnarray*}
 u & = & \gamma_1 \gamma_2 \cdots \gamma_p \gamma'_{p+1}  \\ 
 x & = & \gamma''_{p+1} \gamma_{p+2} \cdots \gamma_{t} 
\end{eqnarray*}
where $\gamma'_{p+1} \gamma''_{p+1} = \gamma_{p+1}$.
Similarly, we can write
\begin{eqnarray*}
 x & = & \eta_1 \eta_2 \cdots \eta_{t-p-1} \eta'_{t-p} \\
 u & = & \eta''_{t-p} \eta_{t-p+1} \cdots \eta_{t} 
\end{eqnarray*}
where $\eta'_{t-p} \eta''_{t-p} = \eta_{t-p}$.

Thus, we have that $\gamma_1 \gamma_2 \cdots \gamma_p \gamma'_{p+1} = \eta''_{t-p} \eta_{t-p+1} \cdots \eta_{t}$. Write $\eta_t = \eta'_t \eta''_t$ where $|\eta''_t| = |\gamma'_{p+1}|$. Similarly,
write $\gamma_p = \gamma'_p \gamma''_p$ where $|\gamma''_p| = |\eta'_t|$. Then we have that $\eta_t = \gamma''_p \gamma'_{p+1}$ and so certainly their images under $\Psi$ are the same. 
As all blocks of $ux$ and $xu$ have the same image, we can therefore conclude that 
$\Psi(\gamma_{p+1}) = \Psi(\gamma''_p \gamma'_{p+1})$.  But clearly $\Psi(\gamma_{p+1}) = \Psi(\gamma'_{p+1}) + \Psi(\gamma''_{p+1})$. Therefore, we get that
$\Psi(\gamma''_{p+1}) = \Psi(\gamma''_{p})$.  Finally, we have that $\Psi(\gamma_p) = \Psi(\gamma_{p+1})$ gives that $\Psi(\gamma'_p) = \Psi(\gamma'_{p+1})$.  Continuing in this way, we
can factorize each $\gamma_i$ into $\gamma'_i$ and $\gamma''_i$ so that all $\gamma'_i$ have the same image under $\Psi$, and separately, all the $\gamma''_i$ have the same image under $\Psi$.

Thus, let $t = r$, $s =p$ and $\alpha_i = \gamma'_i$ and $\beta_i = \gamma''_i$ for all $1 \le i \le r$. Then we get that $u = \alpha_1 \beta_1 \cdots \alpha_s \beta_s \alpha_{s+1}$ and
$x = \beta_s \alpha_{s+1} \beta_{s+1} \cdots \alpha_{r} \beta{r}$. We can then verify that the remaining conditions of the lemma hold using these definitions of $\alpha_i,\beta_i$.
\end{proof}

\begin{example}
If $x = abca$ and $u = cbabc$ then $xu \sim_3 ux$ (which was shown in 
Example~\ref{example_code_m1}). Note that $x$ and $u$ have different lengths and thus cannot 
share an A-primitive root.

\end{example}
The case where both $x$ and $u$ have A-primitive roots of length $n$ is of particular interest:

\begin{corollary}\label{cor_code_1}
Let $u,x \in \Sigma^*$ with $ux \sim_n xu$. If $u$ has an A-primitive root of length $n$, then $x$ does as well, and these A-roots are the same.
\end{corollary}

Corollary~\ref{cor_code_1} is analogous to the second Lyndon-Sch\"utzenberger theorem
(see e.g., Lothaire \cite{L97} or Shyr \cite{S01})
which can be interpreted (in part) as $ux = xu$ if and only if $x$ and $u$ both have
the same primitive root.

We note that the conditions of $\sim_n$ cannot be weakened to allow not all of the subwords of both $u$ and $x$ to have identical images under $\Psi$ and have Theorem~\ref{theorem_code_0} holds, 
as the following example demonstrates:
\begin{example}
Let $\simeq_n$ be the binary relation defined on words by $u \simeq_n x$ if we can write $u = \alpha_1 \alpha_2 \cdots \alpha_m$ and $x = \beta_1 \beta_2 \cdots \beta_m$ where
\begin{enumerate}
\item[(a)] for all $1 \le i \le m$, $|\alpha_i| = |\beta_i| = n$. 
\item[(b)] for all $1 \le i \le m$, $\Psi(\alpha_i) = \Psi(\beta_i)$.
\end{enumerate}
Thus, only parallel subwords of length $n$ are required to be permutations of one another in this definition. But note that if $x = a$ and $u = baa$ then $abaa \simeq_2 baaa$. 
Note that no
factorization of $x$ and $u$ of the form of Theorem~\ref{theorem_code_0} can exist, 
as $u$ cannot be factored as $u =\beta u' \beta$ for any nonempty word $\beta$.
\end{example}

\section{Conclusions}

We have studied the formal language theoretic and combinatorial properties of 
Abelian primitive words.  Unlike classical primitive words, the number of 
Abelian primitive words is a nontrivial combinatorial problem. On the other
hand, we show that the set of Abelian primitive words are not context-free, unlike
the long-standing open problem for primitive words.  Future research problems 
include an exact enumeration of the number of Abelian primitive words of length
$n$. 

\bibliographystyle{plain}

\end{document}